# Abstract Certification of Global Non-Interference in Rewriting Logic[*]

Mauricio Alba-Castro[1,2], María Alpuente[1], and Santiago Escobar[1]

[1] ELP-DSIC, U. Politécnica de Valencia, Spain. {alpuente,sescobar}@dsic.upv.es
[2] U. Autónoma de Manizales, Colombia. malba@autonoma.edu.co

**Abstract.** *Non–interference* is a semantic program property that assigns confidentiality levels to data objects and prevents illicit information flows from occurring from high to low security levels. In this paper, we present a novel security model for global non–interference which approximates non–interference as a safety property. We also propose a certification technique for global non-interference of complete Java classes based on rewriting logic, a very general *logical* and *semantic framework* that is efficiently implemented in the high-level programming language Maude. Starting from an existing Java semantics specification written in Maude, we develop an extended, information–flow Java semantics that allows us to correctly observe global non-interference policies. In order to achieve a finite state transition system, we develop an abstract Java semantics that we use for secure and effective non-interference Java analysis. The analysis produces certificates that are independently checkable and are small enough to be used in practice.

## 1 Introduction

Confidentiality is a property by which information that is related to an entity or party is not made available or disclosed to unauthorized individuals, entities, or processes. One way to protect confidential data is by establishing an *access control policy* [12] that restricts the access to objects depending on the identity or the role performed by the user, meaning that some privilege is required to access confidential data. A user might establish an access control policy by stipulating that no data that is visible to other users be affected by confidential data. Such a policy allows programs to manipulate and modify confidential data as long as the observable data generated by those programs do not improperly reveal information about the confidential data. A security policy of this sort is called a *non-interference policy* [18] because confidential data should not interfere with publicly observable data. Thus, ensuring that a program adheres to a non-interference policy means analyzing how information flows within the program. The mechanism for transfering information through a computing system is called a *channel*. Variable updating, parameter passing, value return, file reading and writing, and network communication are channels. Channels that use a mechanism that is not designed for information communication are called *covert channels* [38]. There are covert channels such as the control structure of a program, termination, timing, exceptions, and resource exhaustion channels. The information flow that occurs through channels is called *explicit flow* [18] because it does not depend on the specific information that flows. The information flow that occurs through the control structure of a program (conditionals, loops, breaks, and exceptions) is called an *implicit flow* [18] because it depends on the value of the condition that guards the control structure. In this paper, we are interested in both explicit and implicit flows for non-interference analysis of deterministic Java programs. However, we do not consider covert channels such as termination, timing, exceptions, and resource exhaustion channels, i.e., releasing information through termination or non termination of a computation, through the time at which an action occurs, or by the exhaustion of a finite shared resource such as memory.

[*] This work has been partially supported by the EU (FEDER) and the Spanish MEC/MICINN under grant TIN 2007-68093-C02-02.

In [1,2], we proposed an abstract methodology for certifying safety properties of Java source code. It is based on *Rewriting logic* (RWL) and is implemented in Maude [14], which is a high-performance language that implements RWL [32]. In [1], we considered integer arithmetic properties that we analyzed as a safety property, whereas in [2] we dealt with (local) non–interference of Java methods. Non-interference is usually defined as a hyperproperty [13], i.e., a property defined on a set of sets of traces, and cannot be established by simply checking a (safety) property on a set of runs (essentially, no single run of a system can violate non-interference). However, we are able to analyze non-interference by observing a stronger property which can be checked as a safety[3] property using an instrumented flow sensitive semantics.

The methodology of [1,2] is as follows. Consider a Java program together with a specification of the Java semantics. The Java program is a concrete expression (i.e., term) that represents the initial state of the Java interpreter running the considered Java program. The Java semantics is a specification in Maude. Given a safety property (i.e., a system property that is defined in terms of certain events that do not happen [30]), the unreachability of the system states that denote the events that should never occur allows us to infer the desired safety property. Unreachability analysis is performed by using the standard Maude (breadth–first) search command, which explores the entire state space of the program from an initial system state. In the case where the unreachability test succeeds, the corresponding rewriting proofs that demonstrate that those states cannot be reached are delivered as the expected outcome certificate. Very often the unreachability test does not succeed because there is an infinite search space; thus, we achieve a finite search space by using abstraction [15]. In our methodology, certificates are encoded as (abstract) rewriting sequences that (together with an encoding of the abstraction in Maude) can be checked by standard reduction. Our methodology is an instance of Proof–carrying code (PCC), a mechanism originated by Necula [36] for ensuring the secure behavior of programs.

---

[3] There are other approaches for proving non–interference as a safety property, which use self-composition [17,6], or flow sensitive security types [27].

This article provides a comprehensive and full-fledged formulation of the abstract non–interference certification methodology of [2]. In that work, we focused on the methodology as well as the PCC and rewriting-based particulars of our approach with a specific emphasis on practicality and good performance. This paper, however, formalizes more foundational semantic security aspects, namely: (i) the characterization of non-interference as a safety property on extended Java computations; (ii) the conditions required by Java programs in order to ensure the correctness of our methodology; (iii) the observational capabilities of an attacker; and (iv) the soundness of our abstract non-interference analysis technique. In our previous work [2], we analyzed (local) non–interference of Java functional methods (i.e. methods that return values). However, in this paper, we are able to analyze entire Java programs, and thus, we consider *global non-interference*.

This paper is organized as follows. In Section 2, we recall the notion of non–intereference and describe a mechanism to specify non-interference policies in JML. In Section 3, we recall the specification of the Java semantics in rewriting logic. In Section 4, we extend this semantics to handle confidential information and formulate a non–interference certification methodology that is based on the unreachability of undesired states in the extended semantics. In other words, by using the extended, information-flow Java semantics, we are able to correctly observe global non–interference policies by checking a stronger safety property that, in our framework, implies non-interference. In Section 5, we develop an approximation of the extended Java semantics that produces a finite search space for any input Java program. By using this abstract semantics (which we implement as a source-to-source transformation of the extended semantics in Maude) we formulate our non-interference analysis and prove its soundness. We include some experiments in Section 6. A thorough discussion of related work is presented in Section 7. Finally, Section 8 presents our conclusions.

## 2 Non–interference

A non-interference policy establishes a confidentiality level for each source program variable of primitive datatypes. It guarantees that actual values of variables with a higher confidentiality level do not influence the output of a variable with a lower confidentiality level during program execution [18,26,38,9,43,19]. It is implicitly assumed that constants that appear in a program always have the lowest confidentiality level (i.e., the considered program is authorized to access secret data, but it does not contain secret data in its code).

A non-interference policy can be represented by a *partially ordered set* $\langle Labels, \leq \rangle$ and a labeling function $Labeling : Var \rightarrow Labels$, where $Labels$ is the finite set of confidentiality levels, $\leq$ is a partial order between confidentiality levels, and $Var$ is the set of source program variables [42,5,27]. There are usually two confidentiality levels: $Labels = \{\texttt{Low}, \texttt{High}\}$. These represent pub-lic non-secret data (low confidentiality) and secret data (high confidentiality), respectively. $\langle Labels, \leq \rangle$ forms a lattice where `Low` is the greatest lower bound or *bottom* element ($\bot$), `High` is the least upper bound or *top* element ($\top$), and `Low` $<$ `High`. The *join* operator ($\sqcup$) is defined as `Low` $\sqcup$ `Low` $=$ `Low`; otherwise, $X \sqcup Y =$ `High`. Enforcing non-interference means that the values of `High`-labeled source variables cannot flow to `Low`-labeled source variables, whereas the values of `Low`-labeled source variables can flow to `High`-labeled source variables. The attacker model for global non–interference that we formalize below assumes that the attacker is passive and can only see the `Low`-labeled source variables of the Java program at the initial and final states and not at the intermediate states. Our methodology can certify programs that have temporal breaches and are still non-interferent.

In order to express confidentiality policies, we use the *Java modeling language* JML [29], which is a property specification language for Java modules. As an interface specification language, JML can describe the names and static information found in Java declarations of Java modules with preconditions (`requires` clauses), postconditions (`ensures` clauses), and assert statements (`assert` clauses), all of which express first–order logic statements. As a behavior specification language, JML can describe how the module behaves when assertions are intermixed within the Java source code. The text of an annotation can either be in one line after the `//@` marker, or in many lines enclosed between the markers `/*@` and `@*/`. They are ignored by traditional compilers. The initial confidentiality level of a variable in a Java program is written with the word `setLabel` as a JML annotation (e.g. `setLabel(var, High)`). The confidentiality label of program variables is `Low` if nothing is specified (i.e., program variables are public by default). We do not need to specify the label of either the formal parameters or local variables because they can be inferred from the confidentiality labels of other program variables if they are properly initialized. These JML annotations, together with the default assumption, define the labeling function of the non–interference policy.

*Example 1.* Consider the following Java program borrowed from [17] that models a bank account and the initial state given by the execution of the function `main`:

```
public class Account { int balance; //@ setLabel(balance, High);
  public boolean extraService;
  public Account() { balance = 0; extraService = false; }
  public void writeBalance(int amount) { balance = amount;
    if (balance>=10000) extraService=true; else extraService=false; }
  private int readBalance() {return balance;}
  public boolean readExtra() {return extraService;}
}
class System { static Account a = new Account();
  public static void main(String[] args) {
    int initbalance; //@ setLabel(initbalance, High);
    initbalance = Integer.parseInt(args[0]);
    a.writeBalance(initbalance); System.out.println(readExtra());  }}
```

This non-interference policy specifies that the object field `balance` of the global object `a` and the initialization parameter `initbalance` (i.e., `args[0]`) hold secret data. This program is insecure w.r.t. this policy since an observer with low access rights can obtain partial information about the variable `balance` via an observation of the non–secret variable `extraService`.

We assume a fixed Java program $P_{\text{Java}}$. $Vars(P_{\text{Java}})$ denotes the set of `static` source variables that may be initialized by the `main` function call. We denote the set of *Low* program variables as $Low(P_{\text{Java}}) = \{var \in Vars(P_{\text{Java}}) \mid Labeling(var) = \texttt{Low}\}$. A program state $St$ is a set of value assignments to program variables. Given $var \in Vars(P_{\text{Java}})$ and a state $St$, $St[var]$ denotes the value of variable $var$ in $St$. We model a *Java program* $P_{\text{Java}}$ as a state transition system between pairs $\langle P, St \rangle$, where $P$ is the current, still-to-be-executed part of the Java program $P_{\text{Java}}$ and $St$ represents the current program state. $\langle P_{\text{Java}}, St_0 \rangle$ denotes the initial *configuration* of standard program execution and $\langle \checkmark, St \rangle$ denotes a final *configuration*, where $\checkmark$ stands for the empty program. Note that we assume that every Java program properly terminates for each set of input data (i.e., we do not consider non-terminating programs, deadlocks, or runtime errors). We also assume deterministic Java programs, without threads or exceptions. $\mapsto_{\text{Java}}$ is the transition relation that describes any possible one-step transition between any two Java program states. An *execution* (or trace) of $P_{\text{Java}}$ is a sequence $\langle P_{\text{Java}}, St_0 \rangle \mapsto_{\text{Java}} \cdots \langle P_i, St_i \rangle \mapsto_{\text{Java}} \cdots \mapsto_{\text{Java}} \langle \checkmark, St_n \rangle$, which is simply denoted by $\langle P_{\text{Java}}, St_0 \rangle \mapsto^*_{\text{Java}} \langle \checkmark, S_n \rangle$ if the intermediate states are irrelevant. We can also abbreviate $\langle \checkmark, S_n \rangle$ by $\langle S_n \rangle$.

We define program non–interference by using an equivalence $=_{Low}$ relationship between states [38,42,5,31]. Roughly speaking, non-interference establishes that any two terminating runs of a program that start from indistinguishable initial states produce indistinguishable final states.

**Definition 1 (State equality [38]).** *Given a Java program $P_{\text{Java}}$, two states $St_1$ and $St_2$ for $P_{\text{Java}}$ are* indistinguishable *at the confidentiality level* `Low`*, written $St_1 =_{Low} St_2$, if for all $var \in Low(P_{\text{Java}}), St_1[var] = St_2[var]$.*

What the attacker can see from a final state is determined by a relation $\approx_{\texttt{Low}}$. Two executions of a program $P_{\text{Java}}$ are related by $\approx_{\texttt{Low}}$ if they are indistinguishable to the attacker [38]. The notion of non–interference is therefore parametric on $\approx_{\texttt{Low}}$. A program is non–interferent if, whenever different initial program states are indistinguishable at level `Low`, this implies that the corresponding final states are also indistinguishable at level `Low`.

**Definition 2 (Non–interference [38]).** *A Java program $P_{\text{Java}}$ is non–interferent if for every pair of different program initial states $St_1$ and $St_2$, and for their corresponding final program states $St'_1$, $St'_2$ such that $\langle P_{\text{Java}}, St_1 \rangle \mapsto^*_{\text{Java}} \langle St'_1 \rangle$ and $\langle P_{\text{Java}}, St_2 \rangle \mapsto^*_{\text{Java}} \langle St'_2 \rangle$, we have that $St_1 =_{Low} St_2$ implies $St'_1 \approx_{Low} St'_2$.*

In this paper, we follow the standard approach in the literature that considers $St \approx_{Low} St'$ iff $St =_{Low} St'$. Then, the non–interference condition of Definition 2 is understood as the lack of any *strong dependence* [38] of `Low`-labeled variables on any of the `High`-labeled variables.

## 3 The Rewriting Logic Semantics of Java

In the following, we briefly recall the rewriting logic semantics of Java that was originally given in [21] and also used by the JavaFAN verification tool [20,22]. We refer the reader to [33] for further technical details on rewriting logic semantics.

In [21], a sufficiently large subset of full Java 1.4 language is specified in Maude, including inheritance, polymorphism, object references, multithreading, and dynamic object allocation. However, Java native methods and many of the available Java built–in libraries are not supported. The specification of Java operational semantics is a rewrite theory: a triple $\mathcal{R}_{Java} = (\Sigma_{Java}, E_{Java}, R_{Java})$ where $\Sigma_{Java}$ is an order–sorted *signature*; $E_{Java} = \Delta_{Java} \uplus B_{Java}$ is a set of $\Sigma_{Java}$–equational *axioms* where $B_{Java}$ are algebraic axioms such as associativity, commutativity and unity, and $\Delta_{Java}$ is a set of terminating and confluent (modulo $B_{Java}$) equations. Finally, $R_{Java}$ is a set of $\Sigma_{Java}$–rewrite rules that are not required to be confluent nor terminating.

Intuitively, the sorts and function symbols in $\Sigma_{Java}$ describe the static structure of the Java program state space as an algebraic data type; the equations in $\Delta_{Java}$ describe the operational semantics of its deterministic features; and the rules in $R_{Java}$ describe its concurrent features. Following the rewriting logic framework [41,32], we denote by $u \to^r_{Java} v$ the fact that the concrete terms $u, v$ (which denote Java program states) are rewritten (at the top position, see [21]) by using $r$, which is either a rule in $R_{Java}$ or an equation in $\Delta_{Java}$ (both of which are applied modulo $B_{Java}$). We simply write $u \to_{Java} v$ when the applied rule or equation is irrelevant. We denote by $\to^*_{Java}$ the extension of $\to_{Java}$ to multiple rewrite steps (i.e., $u \to^*_{Java} v$ if there exist $u_1, \ldots, u_k$ such that $u \to_{Java} u_1 \to_{Java} u_2 \cdots u_k \to_{Java} v$).

The rewrite theory $\mathcal{R}_{Java}$ is defined on terms of a concrete sort `State`, with the main state attributes (represented by means of constructor symbols of the algebraic type `State`) such as `fstack` for handling function calls, `lstack` for handling loops, `env` for assignments of variables to memory locations, and `store` for assignments of memory locations to their actual values. They define an algebraic structure that is parametric w.r.t. a generic sort `Value` that defines all the possible values returned by Java functions or stored in the memory. For instance, the `int` and `bool` constructor symbols describe Java integer and boolean values and are defined in Maude as "`op int : Int → Value .`" and "`op bool : Bool → Value .`", where `Int` and `Bool` are the internal built–in Maude sorts that define integer and boolean data types. Intuitively, equations in $\Delta_{Java}$ and rules in $R_{Java}$ are used to specify the changes to the program state (i.e., the changes to the memory, input/output, etc). Since we consider only deterministic Java programs, our specification of the Java semantics in rewriting logic contains only equations

```
eq k((E > E') -> K) = k((E, E') -> > -> K) .   ---Evaluate arguments
eq k((int(I), int(I')) -> > -> K) = k(bool(I > I') -> K) .  ---Resolve
```

**Fig. 1.** Continuation-based equations for the Java greater-than operator on integers

```
---First obtain location in store from variable name
eq k(Var -> K) env([Var, Loc] Env) = k(#(Loc) -> K) env([Var, Loc] Env) .
---Then obtain value stored in this location
eq k(#(Loc) -> K) store([Loc,Value] Store)
 = k(Value -> K) store([Loc,Value] Store) .
```

**Fig. 2.** Continuation-based equations for variable content retrieval

```
---Obtain variable location and evaluate expression
eq k(Var = E -> K) env([Var, Loc] Env)
 = k(E -> =(Loc) -> K) env([Var, Loc] Env) .
---Once the expression is computed, assign to location
eq k(Val -> =(Loc) -> K) = k([Val -> Loc] -> (Val -> K)) .
---General procedure to update the memory
eq k([Val -> Loc] -> K) store([Loc,Val'] ST) = k(K) store([Loc,Val] ST) .
```

**Fig. 3.** Continuation-based equations for the Java assignment operator

and no rules. The reader can find a RWL specification of the semantics of a programming language with threads in [33,1,2].

The semantics of Java is defined in a *continuation-based style* [33] and specified in Maude itself. Continuations maintain the control context, which explicitly specifies the next steps to be performed. The sequence of actions that still need to be executed are stacked. We use letters K, K' to denote continuation variables, letters E, E' to denote expressions to be evaluated, and Val, Val' to denote values (i.e., the result of evaluating an expression). Once the expression $e$ on the top of a continuation ($e$ -> $k$) is evaluated, its result will be passed on to the remaining continuation $k$. For instance, in Figure 1, the Java greater-than operation on Java integers is specified by using continuations, where k is the constructor symbol used to denote a continuation, -> is the constructor symbol used to concatenate continuations, bool is the constructor symbol used to denote a Java boolean data, and int is the constructor symbol used to denote a Java integer number.

One important aspect of the semantics is the handling of Java variables. In Figure 2, we show how the contents of a Java variable are retrieved from the store (or memory) in the Java state. The semantics of the assignment operator for the Java variables is specified in Figure 3. The if-then-else statement is shown in Figure 4. The semantics of while statements (loops) is specified in Figure 5, where the term while E S denotes the Java iteration statement, the term while(E, S) denotes both the while continuation and the while statement that is expressed in terms of the if(S, S') continuation, and lstack denotes a stack of loops currently being executed, which is needed for a proper control of the

```
--- Evaluates boolean expression keeping the then and else statements
eq k((if E S else S') -> K) = k(E -> (if(S, S') -> K)) .
eq k(bool(true) -> (if(S, S') -> K)) = k(S -> K) .
eq k(bool(false) -> (if(S, S') -> K)) = k(S' -> K) .
```

**Fig. 4.** Continuation-based equations for if-then-else statement

```
--- Stack loop and transform while expression into while continuation
eq k((while E S) -> K) lstack(Lstack)
 = k(while(E,S) -> popLStack -> K) lstack(while(E,S) -> K, Lstack) .
--- A while continuation is transformed into an if-then-else
eq k(while(E,S) -> K) = k(E -> if(S while ( E , S ),{}) -> K) .
--- Add semantics for popLStack
eq k(popLStack -> K) lstack(LItem,Lstack) = k(K) lstack(Lstack) .
```

**Fig. 5.** Continuation-based equations for while statement

```
--- The state is restored from the loop stack
eq k(break -> K) lstack(while(E,S) -> K', Lstack) = k(K') lstack(Lstack) .
```

**Fig. 6.** Continuation-based equations for while break statement

Java `break` statement. Figure 6 shows the semantic specification of the `break` statement, that simply pops the stack of loops. This is important, since it can also abruptly change the information flow. Method calls are not shown in this paper; their semantics is simply defined by eager evaluation of all arguments of the method (whose values are stored in new memory locations) and by creating a new local environment that contains location assignments for formal method parameters and local variables. Due to space limitations we do not discuss heap manipulation here. We refer the reader to [33] for further details.

The following example illustrates the mechanization of the Java semantics.

*Example 2.* Consider again the Java program of Example 1 and two program executions, respectively fed with 5000 and 10000 for the initialization parameter `initbalance`. Note that the corresponding initial states are indistinguishable at the Low confidentiality level (e.g. the only Low-labeled variable, `extraService`, is set to `false` in both of them). The Maude command `search` provides built–in breadth-first search. We ask for the final Java program state of each execution trace (actually, in order to visualize the results, we show the output of `println` Java instructions). The Maude terms `EX1-MAUDE` and `EX2-MAUDE` stand for the Java program with the corresponding initial call (for input value 5000 and 10000, respectively), which are compiled into a Maude expression by using a suitable Java wrapper[4]:

```
search in PGM-SEMANTICS :
java((preprocess(EX1-MAUDE) noType . 'main < new string [i(0)] > noVal))
=>! JO:Output .
```

---

[4] Available at http://fsl.cs.uiuc.edu/index.php/Rewriting_Logic_Semantics_of_Java.

```
Solution 1 JO:Output --> pl(bool(false))
No more solutions.

search in PGM-SEMANTICS :
java((preprocess(EX2-MAUDE) noType . 'main < new string [i(0)] > noVal))
=>! JO:Output .
Solution 1 JO:Output --> pl(bool(true))
No more solutions.
```

If the attacker observes these two final states, she will appreciate the two different values for the variable `extraService`.

## 4 Proving Non–interference by using an Extended Instrumented Semantics

Non–interference is usually understood to be a security property and is therefore defined as a *hyperproperty* [13] (i.e., a property defined on a set of sets of traces). For instance, in Example 2, the verification process for non–interference should check the (possibly infinite) set of (possibly infinite) sets of final states issued from the (possibly infinite) sets of indistinguishable initial configurations. Note that checking the final states issued from `EX1-MAUDE` and `EX2-MAUDE` is just one of the combinations to be analyzed. In contrast, the verification process for a safety property should simply check the traces issuing from the (possibly infinite) set of initial configurations, which is simpler.

In this paper, we prove non-interference as a safety property by instrumenting the Java semantics in order to dynamically keep track of the change of the confidentiality labels of program variables. Intuitively, the semantic instrumentation is defined as follows:

1. Attach a confidentiality label to each memory location; this allows us to observe their confidentiality level at the final execution state.
2. Attach a confidentiality label to the evaluation of program expressions; this allows us to know whether the evaluation of an expression involves high confidentiality data.
3. Associate a confidentiality label to the evaluation of program statements, particularly those involving conditional expressions or guards; this allows us to determine whether the control flow at a given execution point depends on the actual value of high confidential variables. However, this label is not attached to each program statement. Rather it is kept as an extra attribute of a state in the extended Java semantics. This corresponds to the notion of a *context label* being updated after each evaluation step in [18,28,27], which is introduced in the following example.

*Example 3.* Consider the following Java[5] program `TestClass` that is borrowed from [43]. We endow it with the attached non-interference policy:

---

[5] We omit the semantics of some Java operators such as `_++`, `++_`, and `_+=_`, since they can be defined in terms of addition (`_+_`) and assignment (`_=_`), as usual [33].

```
public class Testclass { static int low=0, high; //@ setLabel(high, High);
  public static void main(String[] args) {
    high = Integer.parseInt(args[0]); while (high > 0) {high--;low++;} }}
```

Here there is an an illicit and implicit information flow from the `High`-labeled source variable `high` to the `Low`-labeled source variable `low`. For instance, when the variable `high` contains the value `0` or `1`, the variable `low` is assigned the value `0` and `1`, respectively. This implicit flow would be detected using the context label, which is set to `High` after evaluating the expression `high>0`, and which forces variable `low` to be set to `High` independently of the confidentiality level of the expression `low++`.

In contrast to [2] where local non-interference was studied, here we consider global non-interference (i.e., we are able to ensure a non-interference policy at the final state of the whole Java program execution, which contains several methods, classes, and function calls). This important improvement in the verification power (which has been hardly explored in the related literature) requires the following two modifications to the non-interference analysis of [2]. These changes avoid the difficult (or costly) process of tracing the current confidentiality label of a memory location back to the point where this location was created.

1. We introduce an additional confidentiality label (`Low` $\gg$ `High`), which allow us to represent not only the current confidentiality label of a memory location but also to keep track, at a global level, of hazardous transitions from an initial confidentiality label `Low` to `High`. Similarly, we introduce the confidentiality label (`High` $\gg$ `Low`), in order to avoid false positives where a `High`–labeled variable is updated with the value of a `Low`–labeled expression and then updated again with the value of a `High`–labeled expression.
2. In [2], we used the context label only when updating the value of a variable in memory, as in [28,43,27,24], and when returning values as in [24]. In this paper, we use the context label during expression evaluation, as in [5].

We describe the information-flow extended version of the rewriting logic semantics of Java by the rewrite theory $\mathcal{R}_{\text{Java}^{\text{E}}} = (\Sigma_{\text{Java}^{\text{E}}}, E_{\text{Java}^{\text{E}}}, R_{\text{Java}^{\text{E}}})$, $E_{\text{Java}^{\text{E}}} = \Delta_{\text{Java}^{\text{E}}} \uplus B_{\text{Java}^{\text{E}}}$ and its corresponding $\rightarrow_{\text{Java}^{\text{E}}}$ rewriting relation. In the new semantics, program data not only consist of standard concrete values but each value is decorated with its corresponding confidentiality label. Formally, we consider the label change $LabelChange = \{\text{Low} \gg \text{High}, \text{High} \gg \text{Low}\}$ so that the domain of program variables in the extended semantics is $Value \times (Labels \cup LabelChange)$. We write `<Value,LValue>` for a pair consisting of a concrete value and its corresponding confidentiality label in $Labels \cup LabelChange$.

Thanks to the modularity of the rewriting logic approach to formalizing program semantics [21], our changes to the semantics of Section 3 are incremental and minimal. As Figures 7 and Figure 8 show, the evaluation of constants and variables now uses the context label. As Figure 9 shows, the assignment computes the new confidentiality label in terms of the previous label at the memory location, namely `NewVal = LVal' >>> LVal`. The new operator $\ggg$ is defined in Figure 10.

```
            eq k(i(I) -> K) lenv(CL) = k(<int(I),CL> -> K) lenv(CL) .
            eq k(b(B) -> K) lenv(CL) = k(<bool(B),CL> -> K) lenv(CL) .
```

**Fig. 7.** Extended equations for extended constant evaluation

```
---First obtain location in store from variable name
eq k(Var -> K) env([Var, Loc] Env) = ... .
---Then obtain value stored in this location
eq k(#(Loc) -> K) store([Loc,<Val,LVal>] Store) lenv(CL)
 = k(<Val,LVal join CL> -> K) store([Loc,<Val,LVal>] Store) lenv(CL) .
```

**Fig. 8.** Extended equations for variable content retrieval

```
       ---Obtain variable location and evaluate expression
       eq k(Var = E -> K) env([Var, Loc] Env) = ... .
       ---Once the expression is computed, assign to location
       eq k(<Val,LVal> -> =(L) -> K)
        = k([<Val,LVal> -> L] -> (<Val,LVal> -> K )) .
       ---General procedure to update the memory
       eq k([<Val,LVal> -> Loc] -> K) store([Loc,<Val',LVal'>] ST)
        = k(K) store([Loc,LVal' >>> LVal] ST) .
```

**Fig. 9.** Extended equations for the Java assignment operator

| Previously Stored Label | $\ggg$ | New Label | = | New Stored Label |
|---|---|---|---|---|
| L | $\ggg$ | L | = | L |
| Low | $\ggg$ | High | = | Low $\gg$ High |
| High | $\ggg$ | Low | = | High $\gg$ Low |
| $L_1 \gg L_2$ | $\ggg$ | $L_1$ | = | $L_1$ |
| $L_1 \gg L_2$ | $\ggg$ | $L_2$ | = | $L_1 \gg L_2$ |

**Fig. 10.** Updating memory locations

The context label can only change due to a conditional control flow statements. According to [18,5,28,27], the evaluation of its boolean guards returns a confidentiality level that is associated to the resulting `true` or `false` value and, possibly, a modified context label. The extended semantic equations for the if-then-else of Figure 4 need some slight revision, which is motivated by the following example.

*Example 4.* Consider the following Java method, where the value computed for the variable `low` does not actually depend on the value of the high confidentiality variable `high` (which only affects the temporal variable `aux`). This program does fulfill the non-interference policy at the final state, which can be proved by using our non-interference verification methodology.

```
class Testclass { static int low=0, high; //@ setLabel(high, High);
  public static void main(String[] args) {
     high = Integer.parseInt(args[0]);
     int aux=0; if (high > 2) aux = 1; else aux = 0; low = 0; } }
```

```
--- Evaluates boolean expression keeping the then and else statements
ceq k((if E S else S') -> K) lenv(CL)
 = k(E -> (if(S, S') -> restoreLEnv(CL) -> K)) lenv(CL)
 if not break-or-continue(S) and not break-or-continue(S') .
ceq k((if E S else S') -> K) lenv(CL) = k(E -> (if(S, S') -> K)) lenv(CL)
 if break-or-continue(S) or break-or-continue(S') .
eq k(<bool(true),LVal> -> (if(S, S') -> K)) lenv(CL)
 = k(S -> K) lenv(CL join LVal) .
eq k(<bool(false),LVal> -> (if(S, S') -> K)) lenv(CL)
 = k(S' -> K) lenv(CL join LVal) .
--- New equation to restore previous context label
eq k(restoreLEnv(CL) -> K) lenv(CL') = k(K) lenv(CL) .
```

**Fig. 11.** Extended equations for the if-then-else

In order to avoid false positives during the evaluation of conditional statements, we dynamically restore the previous context label after its execution. The extended semantics equations for the if-then-else are shown in Figure 11, where a new continuation symbol `restoreLEnv` is used to restore the previous confidentiality label. However, restoring the previous context label has to be carefully considered in the presence of `break` or `continue` statements within a loop, since they can abruptly change the information flow as shown in the following example.

*Example 5.* Consider a variation of Example 3 where the while loop has a bogus guard together with a `break` statement to exit the loop:

```
public class Testclass { static int low=0, high; //@ setLabel(high, High);
 public static void main(String[] args) {high = Integer.parseInt(args[0]);
    int aux=0; while (true) {high--; low++; if (high == 0) break;} } }
```

As in Example 3, when the while loop ends, the variable `low` has the initial value of the variable `high`. Whenever $high \neq 0$, the `break` statement is not executed. In this case, the conditional guard uses `High`-labeled data, and the conditional statement should not restore the previous context label. In other words, the critical component here is not the `break` statement but rather the else branch that does not contain the `break`.

In order to solve this problem, we check in Figure 11 whether either of the two branches of a conditional statement contains a `break` or `continue` statement and no other conditional statement or `while` loop in between. If there is such a statement, `restoreLEnv` is not used. This case was not considered in [43] or in [2], which only considered `break` statements within `High` guarded while loops.

Method invocation propagates the context label without changes as proposed in [28] and, thus, is not shown here. Since while statements were expressed in terms of if-then-else statements, they need a slight extension to introduce the `restorelEnv` continuation (shown in Figure 12). The semantic specification of the `break` statement stays the same as shown in Figure 6: the context label

```
--- Stack loop and transform while expression into while continuation
eq k((while E S) -> K) lstack(Lstack) lenv(CL)
 = k(while(E,S) -> restoreLEnv(CL) -> popLStack -> K)
   lstack(while(E,S) -> K, Lstack) lenv(CL) .
```

**Fig. 12.** Extended equations for while statement

lenv(CL) is not modified and the `restoreLEnv` expression introduced by the while statement is removed.

### 4.1 Proving non-interference as a safety property

Now, we are ready to formulate a novel characterization of non-interference that allows us to check it as a property that is verified for each possible execution trace instead of being verified for each set of indistinguishable execution traces.

**Definition 3 (Strong Non-Interference).** *A Java program $P_{\text{Java}}$ is strongly non–interferent for a given labeling function if for every extended initial state $St_1^E$ and for its corresponding final program state $St_2^E$ given by $\langle P_{\text{Java}}, St_1^E \rangle \mapsto^*_{\text{Java}^E} \langle St_2^E \rangle$, we have that for all $var \in Low(P_{\text{Java}})$, $St_2^E[var] = \langle Val, \text{Low} \rangle$ for a value $Val$.*

Since in our model, a public variable can only have the label Low or the label Low $\gg$ High, this means that in the extended execution of a program that is not strongly non-interferent, the label of at least one program variable is Low $\gg$ High. Given an initial state $St$ and a given labeling function, we denote the corresponding extended state by $St^E$.

**Lemma 1.** *Consider a Java program $P_{\text{Java}}$ and two initial states $St_1$ and $St_2$ such that $St_1 =_{Low} St_2$. Consider the two corresponding final program states $St_1'$ and $St_2'$ given by $\langle P_{\text{Java}}, St_1 \rangle \mapsto^*_{\text{Java}} \langle St_1' \rangle$, $\langle P_{\text{Java}}, St_2 \rangle \mapsto^*_{\text{Java}} \langle St_2' \rangle$. If there exists $var \in Low(P_{\text{Java}})$ such that $St_1'[var] \neq St_2'[var]$, then $\langle P_{\text{Java}}, St_1^E \rangle \mapsto^*_{\text{Java}^E} \langle St^E \rangle$ and $St^E[var] = \langle Val, \text{Low} \gg \text{High} \rangle$ for a value $Val$.*

*Proof.* Consider the two traces $\mathcal{D}_1 : \langle P_{\text{Java}}, St_1 \rangle \mapsto^*_{\text{Java}} \langle St_1' \rangle$ and $\mathcal{D}_2 : \langle P_{\text{Java}}, St_2 \rangle \mapsto^*_{\text{Java}} \langle St_2' \rangle$. Let $\{var_1, \ldots, var_k\} \subseteq Low(P_{\text{Java}})$ be those variables such that $St_1'[var_i] \neq St_2'[var_i]$ for all $1 \leq i \leq k$. Since we assume $k > 0$, then there is at least one of those variables (say $var_1$) and an assignment statement $var_1 = E_1$ that is executed at least once in one of the two traces (say $\mathcal{D}_1$). Let $n$ be the total number of assignments in $\mathcal{D}_1$ to variables $\{var_1, \ldots, var_k\}$. Note that $n$ is finite since execution traces are finite because of the termination assumption. Now, we prove the result by induction on $n$.

1. ($n = 1$) Let us consider the last execution step in $\mathcal{D}_1$ where the assignment $var_1 = E_1$ is executed. Then, it may happen that the assignment $var_1 = E_1$ is also executed in $\mathcal{D}_2$, or not. We consider these two cases separately.

(a) If $var_1 = E_1$ is also executed in $\mathcal{D}_2$, then $St'_1[var_1] \neq St'_2[var_1]$ implies that the values for $E_1$ are different in the two traces. Thus, expression $E_1$ must contain at least one variable $var'$ such that the actual values of $var'$ are different in the two traces when the considered assignments to $var_1$ are executed. Since $St'_1[var'] \neq St'_2[var']$ and $n = 1$, then $var' \notin Low(P_{\text{Java}})$. Therefore $var'$ is a High confidentiality variable, hence it has a High label in our extended semantics. This means that the label Low $\gg$ High is assigned to variable $var_i$ (according to Figure 10) in $\mathcal{D}_1$, and the conclusion follows.

(b) If $var_1 = E_1$ is not executed in $\mathcal{D}_2$, then $St'_1[var_1] \neq St'_2[var_1]$ implies that the execution of this last assignment statement $var_1 = E_1$ in $\mathcal{D}_1$ is conditioned to the result of a boolean expression containing High confidentiality variables that guards a conditional (or while loop) statement so that the assignment is executed in $\mathcal{D}_1$ and not in $\mathcal{D}_2$. Then, the assignment statement $var_1 = E_1$ in $\mathcal{D}_1$ was executed either (i) within the then or else branch of an if-then-else Java statement (recall that while loops are expressed as if-then-else statements), (ii) within the then branch of an if-then Java statement, or (iii) after evaluating a conditional expression within a while loop that includes a break expression. Note that no other case can generate an interference condition. In all three cases, our extended semantics assigns a High label to the boolean guard expression of such a conditional expression and the context label is set to High (according to Figures 11 and 12) before the expression $E_1$ is evaluated in the statement $var_1 = E_1$. Note that in case (iii), the conditional expression propagates the High context label outside itself (according to Figure 6), i.e. the conditional does not restore the previous context label precisely to record that even if sequence $\mathcal{D}_1$ does not execute the break statement, another possible trace (e.g. $\mathcal{D}_2$) can do it. Finally, in all three cases, the expression $E_1$ is evaluated within a High–labeled context and then the label Low $\gg$ High is assigned to variable $var_1$, independently of whether expression $E_1$ manipulates High confidential data or not.

2. ($n > 1$) Let us consider the *last* execution step in $\mathcal{D}_1$ where the assignment $var_i = E_i$ is executed, with $1 \leq i \leq k$. We split into two cases.

    (a) If $var_i = E_i$ is also the last assignment of variables $\{var_1, \ldots, var_k\}$ executed in $\mathcal{D}_2$, then $St'_1[var_i] \neq St'_2[var_i]$ implies that the values for $E_i$ are different in the two traces. Thus, expression $E_i$ must contain at least one variable $var'$ such that the actual values of $var'$ are different in the two traces when the considered assignments to $var_i$ are executed. Then, let us consider whether $var' \in \{var_1, \ldots, var_k\}$ or not. If it is, then by induction hypothesis, we can assume that variable $var'$ has a Low $\gg$ High label since we can replace the execution of the last assignment $var_i = E_i$ by a simple $var_i = cte$ (where $cte$ is a constant) and the program will still be interferent, due to the fact that the assignment to $var'$ occurs before and could not be affected by the last assignement to $var_i$. If $var' \notin \{var_1, \ldots, var_k\}$, then $var'$ is a High confidentiality variable and it has a High label in our extended semantics. In both cases, the label

Low ≫ High is assigned to variable $var_i$ (according to Figure 10), and the conclusion follows.

(b) If $var_i = E_i$ is not the last assignment of variables $\{var_1, \ldots, var_k\}$ executed in $\mathcal{D}_2$, then either there is no such an assignment in $\mathcal{D}_2$ to variables $\{var_1, \ldots, var_k\}$, or the last assignment in $\mathcal{D}_2$ has the form $var_i = E'$, with $E'$ different from $E_i$, or it affects a variable $var''$ that is different from $var_i$. All three cases imply that the execution of the last assignment statement $var_i = E_i$ in $\mathcal{D}_1$ is conditioned to the result of a boolean expression containing High confidentiality variables that guards a conditional (or while loop) statement so that such assignment is executed in $\mathcal{D}_1$ and not in $\mathcal{D}_2$. Then this case is perfectly similar to case (1)(b) above, and the result follows. □

From Lemma 1 we derive that strong non-interference implies non-interference, as given by the following result.

**Theorem 1 (Strong Non-Interference Soundness).** *Given a Java program $P_{\text{Java}}$, if $P_{\text{Java}}$ is strongly non–interferent (Definition 3), then $P_{\text{Java}}$ is non–interferent (Definition 2).*

*Proof.* (By contradiction) Assume that program $P_{\text{Java}}$ is strongly non–interferent and also that $P_{\text{Java}}$ is interferent. Since $P_{\text{Java}}$ is strongly non–interferent, for every extended initial state $St^E$ and for its corresponding final program state $St^{E'}$ given by $\langle P_{\text{Java}}, St^E \rangle \mapsto^*_{\text{Java}^E} \langle St^{E'} \rangle$, we have that for all $var \in Low(P_{\text{Java}})$, $St^{E'}[var] = \langle Val, \text{Low} \rangle$ for a value $Val$. By Lemma 1 and the assumption that $P_{\text{Java}}$ is interferent we have that $St^{E'}[var] = \langle Val, \text{Low} \gg \text{High} \rangle$ for a value $Val$, hence $P_{\text{Java}}$ is not strongly non–interferent, contradicting the hypothesis. □

The following example illustrates the mechanization of our verification methodology.

*Example 6.* Consider again the Java program of Example 1. Now, we compute the final state in the extended Java program execution for EX1-MAUDE (for simplicity we show only the value of variable extraBalance).

```
search in PGM-SEMANTICS-EXTENDED :
java((preprocess(EX1-MAUDE) noType .  'main < new string [i(0)] > noVal))
=>! M:Store .
Solution 1 M:Store --> store([l(6),<bool(false),Low >> High>] ...)
No more solutions.
```

The execution for EX2-MAUDE will also contain the label Low ≫ High for variable extraBalance.

In other words, we transform non-interference into a stronger property which can be effectively checked in the extended semantics. Obviously, we are not able to certify the security of all the programs that are secure, as shown in Example 7.

*Example 7.* Consider the following Java program borrowed from [43].

```
class Testclass { static int low=0, high; //@ setLabel(high, High);
 public static void main(String[] args) {high = Integer.parseInt(args[0]);
     low = high; low = low - high;} }
```

Apparently, there is an explicit flow from variable `high` to variable `low` through the two assignment statements. However for any execution, when program ends, the value of variable `low` is always `0` so that the variable `low` does not depend on the variable `high`. According to Definition 2, the program is non–interferent. However, we give a false positive by using our notion of strong non-interference since the assignment "$low = high$" assigns to the variable `low` a high confidentiality label $Low \gg High$ and the last statement "$low = low - high$" does not revert the label back to `low`.

The program of Example 7 cannot be verified by traditional type inference approaches [42,46,4] either, since they fail to verify (type check) any program with temporary breaches, e.g. Examples 4 and 7 above, whereas Example 4 is effectively verified by using our methodology.

## 5 Approximating Non–interference by using an Abstract Semantics

The extended, instrumented Java semantics defined so far allows us to develop a technique for proving non–interference. However, this technique is still not feasible in general because there are too many possible initial states to consider for the safety property to be checked. In the following, we develop an abstract, rewriting logic Java semantics that allows us to statically analyze global non–interference. Similar to [2], the purpose of the abstract semantics is to correctly approximate the extended computations in a finite way. Given the extended Java semantics, where there are concrete labeled values, we simply get rid of the values in the abstract semantics, and use their confidentiality labels as the abstract values instead.

In the following, we develop an abstract version of the extended rewriting logic semantics of Java developed in Section 4, which we describe by the rewrite theory $\mathcal{R}_{\text{Java}\#} = (\Sigma_{\text{Java}\#}, E_{\text{Java}\#}, R_{\text{Java}\#})$, $E_{\text{Java}\#} = \Delta_{\text{Java}\#} \uplus B_{\text{Java}\#}$ and its corresponding $\rightarrow_{\text{Java}\#}$ rewriting relation. As in Section 4, our approach for the abstract Java semantics consists of modifying the original theory $\mathcal{R}_{\text{Java}^E}$ (taking advantage of its modularity) by abstracting the domain to $Labels \cup LabelChange$ and introducing approximate versions of the Java constructions and operators tailored to this domain.

An *abstract interpretation* (or abstraction) [16] of the program semantics is given by an *upper closure operator* $\alpha : \wp(\mathsf{State}) \rightarrow \wp(\mathsf{State})$, which is *monotonic* (for all $SSt_1, SSt_2 \in \wp(\mathsf{State})$, $SSt_1 \subseteq SSt_2$ implies $\alpha(SSt_1) \subseteq \alpha(SSt_2)$), *idempotent* (for all $SSt \in \wp(\mathsf{State})$, $\alpha(SSt) \subseteq \alpha(\alpha(SSt))$), and *extensive* (for all $SSt \in \wp(\mathsf{State})$, $SSt \subseteq \alpha(SSt)$). In our framework, each Java program state $St \in \mathsf{State}$ is abstracted by its closure $\alpha(\{St\})$. Our abstraction function

```
rl k(LVal -> (if(S,S') -> K)) lenv(CL) => k(S -> K) lenv(CL join LVal) .
rl k(LVal -> (if(S,S') -> K)) lenv(CL) => k(S' -> K) lenv(CL join LVal) .
```

**Fig. 13.** Abstract rules for the if-then-else

$\alpha : \wp(\mathsf{State}^E) \to \wp(\mathsf{State}^E)$ is a simple homomorphic extension to sets of states of the function $2nd : \mathsf{Value} \times (Labels \cup LabelChange) \to (Labels \cup LabelChange)$, meaning that we disregard the actual values of data.

In the abstract Java semantics, several alternative computation steps of $\to_{\mathrm{Java}^E}$ are mimicked by a single abstract computation step of $\to_{\mathrm{Java}\#}$, reflecting the fact that several distinct behaviors are compressed into a single abstract state (i.e. set of states). The instrumentalization of the Java semantics for dealing with a set of states instead of one single state implicitly means too many modifications. Therefore, we adopt a different approach. When several $\to_{\mathrm{Java}^E}$ rewrite steps are mimicked by a single abstract rewriting state leading to an abstract Java state, and those rewrite steps apply different rules or equations, we use concurrency at the Maude level. Despite the fact that our extended Java semantics contains only equations and no rules, the abstract Java semantics does contain rules in $R_{\mathrm{Java}\#}$ to reflect the different possible evolutions of the system.

The abstract semantics is mainly a straightforward extension of the extended semantics. The only difference is that any set of equations that was confluent and terminating in the extended semantics but might become non confluent or non terminating in the abstract semantics is transformed into rules. As a representative example, the abstract rules associated to two of the equations of the extended semantics of the if-then-else statement are shown in Figure 13.

Now, we are ready to formalize the abstract rewriting relation $\to_{\mathrm{Java}\#}$, which intuitively develops the idea of applying only one rule or equation from the concrete Java semantics to an abstract Java state while exploring the different alternatives in a non-deterministic way. By abuse, we denote the abstraction of a rule $\alpha(\{l\}) \to \alpha(\{r\})$ by $\alpha(\{l\} \to \{r\})$. $\mathcal{P}_{\mathrm{Java}}$ denotes the sort of Java programs $P_{\mathrm{Java}}$ (i.e. $P_{\mathrm{Java}} \in \mathcal{P}_{\mathrm{Java}}$).

**Definition 4 (Abstract rewriting).** *We define abstract rewriting* $\to_{\mathrm{Java}\#} \subseteq (\mathcal{P}_{\mathrm{Java}} \times \wp(\mathsf{State}^E)) \times (\mathcal{P}_{\mathrm{Java}} \times \wp(\mathsf{State}^E))$ *by* $\langle P_{\mathrm{Java}_1}, SSt_1 \rangle \to_{\mathrm{Java}\#} \langle P_{\mathrm{Java}_2}, SSt_2 \rangle$ *if* $\exists u \in SSt_1, \exists v \in SSt_2$ *s.t.* $\langle P_{\mathrm{Java}_1}, u \rangle \to_{\mathrm{Java}^E} \langle P_{\mathrm{Java}_2}, v \rangle$.

We denote by $\to_{\mathrm{Java}\#}^*$ the extension of $\to_{\mathrm{Java}\#}$ to multiple rewrite steps.

**Lemma 2.** *If* $\langle P_{\mathrm{Java}}, St_1^E \rangle \to_{\mathrm{Java}^E}^* \langle St_2^E \rangle$, *then there exists* $SSt_3 \in \wp(\mathsf{State}^E)$ *s.t.* $\langle P_{\mathrm{Java}}, \alpha(\{St_1^E\}) \rangle \to_{\mathrm{Java}\#}^* \langle SSt_3 \rangle$ *and* $St_2^E \in SSt_3$.

*Proof.* (Sketch) Our abstraction consists of transforming equations into rules and getting rid of the value component of states. Since the transformation of a set of equations (which are confluent and terminating modulo axioms) into rules preserves the execution traces, and (by the monotonicity, idempotency, and extensivity of the upper closure operator $\alpha$) the removal of the value component of states does not eliminate execution traces either, then the conclusion follows
□

A program is non–interferent for a given labeling function if the abstract values (the confidentiality labels) of the *Low* variables in the final state of an abstract program execution do not have the label $\texttt{Low} \gg \texttt{High}$.

**Theorem 2 (Abstract Non-Interference Soundness).** *Given a Java program $P_{\text{Java}}$, $P_{\text{Java}}$ is non–interferent (Definition 2) if for all $SSt_1 \in \wp(\textsf{State}^E)$ s.t. $\langle P_{\text{Java}}, SSt_1 \rangle \mapsto^*_{\text{Java}\#} \langle SSt_2 \rangle$, for all $St \in SSt_2$, and for all variables $var \in Low(P_{\text{Java}})$, $St[var] = \langle Val, \texttt{Low} \rangle$ for a value $Val$.*

*Proof.* By contradiction. Let us assume that $P_{\text{Java}}$ is not non–interferent, i.e., there exists $St_1^E$ with $\langle P_{\text{Java}}, St_1^E \rangle \mapsto^*_{\text{Java}^E} \langle St_2^E \rangle$ and $var \in Low(P_{\text{Java}})$ s.t. $St_2^E[var] = \langle Val, L \rangle$ for a value $Val$ and $L \neq \texttt{Low}$. Since $\langle P_{\text{Java}}, St_1^E \rangle \mapsto^*_{\text{Java}^E} \langle St_2^E \rangle$, by Lemma 2, there exists $SSt_3 \in \wp(\textsf{State}^E)$ s.t. $\langle P_{\text{Java}}, \alpha(\{St_1^E\}) \rangle \rightarrow^*_{\text{Java}\#} \langle SSt_3 \rangle$ and $St_2^E \in SSt_3$. This contradicts the assumption that for all $St \in SSt_3$, and for all variables $var \in Low(P_{\text{Java}})$, $St[var] = \langle Val', \texttt{Low} \rangle$ for a value $Val'$. □

The following example illustrates the mechanization of the Java non-interference analysis.

*Example 8.* Consider again the Java program of Example 1. By virtue of the abstraction, we consider just one abstract initial state that safely approximates any extended initial state and compute the corresponding abstract final states.

```
search in PGM-SEMANTICS-ABSTRACT :
java((preprocess(EX1-MAUDE) noType . 'main < new string [i(0)] > noVal))
=>! M:Store .
Solution 1 M:Store --> store([l(6),Low >> High] ...)
No more solutions.
```

Due to the transformation of some equations into rules in the abstract semantics, there may be several execution paths but all lead to the same abstract final state.

## 6 Experiments

Our methodology generates a safety certificate which essentially consists of the set of (abstract) rewriting proofs that implicitly describe the program states which can (and cannot) be reached from a given (abstract) initial state, as illustrated in Example 8. Since these proofs correspond to the execution of the abstract Java semantics specification, which is made available to the code consumer, the certificate can be unexpensively checked on the consumer side by any standard rewrite engine by means of a rewriting process that can be very simplified. Actually, it suffices to check that each abstract rewriting step in the certificate is valid and that no rewriting chain has been disregarded, which essentially amounts to using the matching infrastructure available within the rewriting engine. Note that, according to the different treatment of rules and equations in Maude, where only transitions caused by rules create new states in the space state, an extremely reduced certificate can be delivered by just recording the rewrite steps given with the rules, while the rewritings using the equations are omitted.

The abstract certification methodology described here has been implemented in Maude[6]. The prototype system offers a rewriting-based program certification service, which is able to analyze global confidentiality program properties related to non–interference. Our certification tool can generate three types of certificates: (i) the full certificates consist of complete rewriting sequences including all rewrite steps; (ii) the reduced rules certificates only contain the rewrite steps that use rules; and (iii) the reduced labels certificates only record the labels of the used rules.

| Code Examples → <br> Experiment Measures ↓ | 1 | 2 | 3 | 4 | 5 |
|---|---|---|---|---|---|
| Code size in LOC | 27 | 31 | 48 | 80 | 117 |
| Code size in bytes | 869 | 924 | 1981 | 3305 | 3504 |
| Code cyclomatic complexity | 1 | 1 | 4 | 16 | 192 |
| Full Cert. size (Kb) | 1134 | 1251 | 4223 | 10619 | 24176 |
| Red. Rules Cert. size (Kb) | 6.1 | 6.3 | 21.1 | 47.1 | 21.3 |
| Red. Labels Cert. size (Kb) | 1.8 | 1.8 | 2.6 | 3.7 | 5.2 |
| Full Cert. Gen. Time (ms) | 10408 | 23574 | 29482 | 45709 | 84331 |
| Red. Rules Cert. Gen. Time (ms) | 7057 | 7030 | 7527 | 8215 | 9547 |
| Red. Labels Cert. Gen. Time (ms) | 7030 | 6700 | 7190 | 8198 | 9537 |

**Table 1.** Code measures, certificate sizes, and generation times

In Table 1, we analyze three key points for the practicality of our approach: the size and complexity of the program code, the size of the three types of certificates, and the certificate generation times. The running times are given in milliseconds and were averaged over a sufficient number of iterations. We considered three code measures, the code size in LOC (lines of source code), the code size in bytes, and the cyclomatic complexity, which counts the execution paths of a program. The experiments were performed on a laptop with a Pentium M 1.40 GHz processor and 0.5 Gb RAM.

Program 1 consists mainly of a simple non–interferent code example borrowed from [43,28]. The program has been structured into two classes. The first class has one secret variable and one public variable, a constructor method, two get methods, and a method that contains the non–interferent piece of code of [43,28]. The second class is the main class with four method invocations. Similarly, program 2 is a simple non–interferent example borrowed from [27]. It is structured into two classes. Program 3 includes three simple methods in two classes: the non–interferent method included in program 1, an interferent method borrowed from [43,28], and another non–interferent method borrowed from [39]. The main method has a sequence of method invocations such that the last invocation calls a non–interferent method, and thus the entire program is non–interferent. Program 4 includes six simple methods, the three methods included in program 3 and three other interferent methods also borrowed from [43,28], including a method with a `while` loop and a method that calls another method. In this case, the last invoked method as well as the whole example program are non-interferent. Similarly, program 5 includes nine simple methods, the six examples included in program 4 plus three other interferent methods: two interferent variations of the loop example of program 5 and an interferent method with a return statement within a conditional statement. The source code of our benchmarks is provided within the distribution package.

---

[6] The tool is provided with a Web interface written in Java and is publicly available at http://www.dsic.upv.es/users/elp/toolsMaude/GlobalNI.hml.

The experiments are very encouraging since they show that the reduction in size of the certificate is very significant in all cases, with the quotient "Red. Rules Cert. Size/Full Cert. Size" ranging from 0.54% in program 2 to 0.09% in program 5. Note that the biggest reduction occurs for the largest program. When the time employed to generate the full and reduced rules certificates are compared, the reduced certificate generation time vs the full certificate generation time range from 11, 32% to 67.80%. The reduction for the biggest example (program 5) was the largest one (11, 32%). Note that the generation time for the reduced labels certificate were not significantly lower than the reduced rules certificate. These results show that the technique scales up better when reduced certificates are considered.

## 7 Related Work

Goguen and Meseguer [26] formalized non–interference of deterministic and terminating systems as a system hyperproperty [13], i.e., a security property that is defined for pairs of system output traces that are indistinguisable for an observer. In [23], Foccardi and Gorrieri defined a stronger, security–based notion of non–interference that considers pairs of system input/output traces. In contrast to [23], our safety-based notion of strong non–interference only considers secret outputs, similarly to [26].

Barthe et. al [6] develop a methodology to prove non–interference of deterministic terminating programs in an imperative language with loops, conditionals, and mutable data structures (i.e. objects). Their methodology relyies on using Hoare logic and separation logic, and handles non–interference as a safety property by using program self–composition with variable renaming (i.e., they compose a program with a copy of itself without sharing memory positions). Their method can verify non–interference of secure programs with temporary breaches such as "`low = high; low = 2`", whereas imprecise conservative type systems [42,46,4] cannot. Also, their method can deal with Examples 4 and 7, whereas we cannot ensure security for the last example. This proposal is complete and sound, but the criterion is undecidable, and for the best of our knowledge no approximation has yet been implemented.

Existing Java verification tools that use standard JML [29] as a property specification language do not support non–interference certification. Some sophisticated non–interference policies can be expressed by using the JML extensions of the Krakatoa Java verification tool [19]. These JML extensions were developed for Hoare-style assertions regarding program self-composition [6]. This means duplicating the code of the program and makes it necessary to distinguish the same program variables in its two runs. These JML extensions are used to express non–interference pre– and post–conditions, but they do not handle confidentiality labels of program variables explicitly; the method assumes that all the variables annotated with the extended JML assertions called "`ni1`" and "`ni2`", are labeled `Low`. Nevertheless, the confidentiality aspect of non–interference is expressible using the JML specification pattern suggested in [28,43] as an instrument for program verification using the theorem prover PVS. Unfortunately, this proposal abuses notation by identifying the confidentiality levels with the values of program variables, and it does not consider important Java features such as method calls and interruptions (`break`, `return` or `continue` statements) within conditional instructions and iterations. Moreover, a specification pattern for confidentiality cannot be created in all cases, as mentioned in [43]. A flow-sensitive and termination-insensitive analysis for object-oriented programs based on Hoare logic is proposed in Amtoft et. al [3]. This analysis considers pointer aliasing that can leak confidential information. The non–interference property is specified by using independence assertions that are written in JML. In order to compute postconditions, the analysis uses an algorithm that is sound and complete given some assumptions, but it does not generate a program security proof.

Although non–interference has not been considered in current PCC implementations, there are some proposals that are based on type systems for a subset of Java [7], Java bytecode [37,9,8], and simple imperative languages [42,27,11]. None of these use JML to express non–interference policies and none of them have yet been implemented. In [7], a type system is proposed as a basis for deriving a certifying compiler for a subset of Java source code with objects, inheritance, methods and simplified exceptions. JFlow [34] and Jif [35] are security-typed programming languages with support for enforcing information-flow and access control with dynamic label policies, at both compile time and run time. These compilers produce secure Java source code for verified programs. In order to deal with program variables whose confidentiality labels are only known at run time, dynamic labels are introduced. However, the dynamic labels of Jif have not yet been proved to enforce secure information flow [47]. Volpano et al [42] developed an information-flow type system that can be used to check non–interference of programs written in a generic deterministic sequential imperative language, but this system cannot verify safe programs that have temporary breaches. In [27], Hunt and Sands propose a flow sensitive, dynamic type system that has not yet been implemented. It tracks syntactical dependences between program variables in a simple imperative language without objects or function calls. Although we consider only two security levels, our methodology can easily been extended to the multilevels of confidentiality of [27,8]. Moreover, we have shown that our analysis can achieve more precision than traditional, type-based approaches, thanks to the combination of static analysis and dynamic labeling. In [8], Barthe et al. define the first information-flow type system for a sequential JVM-like language with classes, objects, arrays, exceptions and method calls that certifies non–interference in type-checked programs. The soundness was proved by using the theorem prover Coq, and a certified lightweight bytecode verifier for information flow was extracted from the proof.

Wasserrab et. al present in [44] the first machine-checked correctness proof for information-flow control that is based on program dependence graphs using static intraprocedural slicing. The proof is formalized in Isabelle/HOL. The analysis applies to deterministic terminating programs and is flow-sensitive, object-sensitive and context-sensitive. The machine-checked proof was instantiated for a simple imperative language with loops and for a subset of Jinja (a definition of Java bytecode), which must be manually annotated with security labels. This work does not consider method calls, classes, or objects. Bavera and Bonelli [10] present a flow-sensitive type system for verifying non–interference of bytecode, where class fields may have different confidentiality labels for different instance objects. This methodology does not consider method calls and it does not generate checkable proofs. Moreover, as is usually the case in type-based analysis, once the object fields and the variable labels are determined, they remain fixed throughout the analysis. A proposal that deals with dynamic information-flow policies is [40]. This technique is based on runtime tracking of indirect dependencies between program points. While our confidentiality label tracking is also dynamic, our approach is based on static analysis rather than runtime monitoring, similarly to [27,28].

Some proposals also exist for non–interference verification that are based on abstract interpretation [5,46,25,24,45]. However, these proposals do not generate a certificate as an outcome of the verification process, and they do not use JML to express non–interference policies. The idea of first enriching the original semantics of the language by pairing each data value to its security level, and then approximating it by only considering the security level was also proposed in [5,46]. A similar idea is used in [24], where an abstract information–flow sensitive collecting semantics, which is called instruction–level security typing, for programs with dynamic structures is proposed; here input and ouput channels are given security levels, but the variables have no associated security levels. A different notion of abstract non–interference is proposed in [25] that approximates the standard notion of non–interference by making it parametric relative to input/output abstractions. In abstract non–interference, the abstract domains encode the allowed flows that characterize the degree of precision of the knowledge of a potential attacker observing the data. By using classes and class hierachies as abstract domains, Zanardini adopts a different perspective of abstract non–interference for classes in [45], where the abstract value of a concrete object is its class. Two objects (values) are indistinguishable at an abstraction level (class) if the objects belong to the given class or if the given class is a superclass of object classes. An algorithm for checking abstract non–interference of Java classes is proposed that relies on class–based dependencies.

In previous work [2], we dealt with (local) non–interference of function methods regarding explicit inputs by parameter passing and explicit outputs by value returning. The local non–interference policies considered there were required to explicitly establish the confidentiality labels for all method parameters and vari-ables. In this work, however, we consider global non–interference of complete Java classes and we do not need to explicitly state the confidentiality level for all program variables. In [2], we worked directly with an implementation level definition of non–interference; in this work, we provide a general and language-independent characterization as well as a formal and rigorous relation between the approximate properties and the security model. As in [18,5,38,28,27], we take into account implicit information flows by considering the context confidentiality label in expression evaluation (the context label is joined with the confidentiality label of the expression) and also by modifying the context label during the evaluation of guards of conditionals and while loops. Our global policies are very flexible since the security levels of object variables, local variables, and method parameters may change temporarily as in [27,28,5,24,6].

## 8 Conclusion

In this paper, we formalize a framework for automatically certifying global non–interference of Java programs. Our methodology relies on an (abstract) extended semantics for Java written in rewriting logic that can be model–checked in Maude by using Maude's breadth-first search space exploration. In the extended semantics, non-interference becomes a safety property, and we formally demonstrate that the safety property in the extended semantics entails the semantic, non-interference security property in the standard Java semantics. In this work, we provide a general and abstract definition as well as a rigorous link between the approximate properties and the security model that we consider, whereas in our previous work [2], we worked directly with a program-level definition of non–interference. The proposed framework fully accounts for explicit as well as implicit flows, and allows not only the inference of rewriting logic safety proofs but also the checking of existing ones, thus providing support for proof-carrying code. Actually, the steps that the abstract semantics takes are recorded in order to construct a certificate ensuring that the program satisfies the desired property. By turning a potentially infinite labelled state space of a Java program into a finite abstract space, the abstract semantics not only makes the approach feasible, but also greatly reduces the size of the certificates that must be checked on the consumer's end.

The Java operational semantics in rewriting logic that we have used is modular and has 2635 lines of code in 4 files [21]. We have modified less than 20 of the 1527 lines of code in the main file of the original Java semantics. The abstract operational Java semantics was developed as a source–to–source transformation in rewriting logic and consists of 650 lines of extra code. This is equivalent to saying that, in our current system, the *trusted computing base* (TCB)[7] is less than a fourth of the size of the original Java semantics (at least one order of magnitude smaller than the standard rewriting infrastructure, and even much smaller than other PCC systems).

---

[7] The TCB is the part of the code that is used to check if other code can be safely run, and it is assumed to be trusted.

Since our approach is based on a rewriting logic semantics specification of the full Java 1.4 language [33], the methodology developed in this work can be easily extended to cope with exceptions, heaps, and multithreading since they are considered in the Java rewriting logic semantics.